\documentclass[printer]{aa}
\usepackage[varg]{txfonts}
\usepackage{natbib,twoopt}
\bibpunct{(}{)}{;}{a}{}{,}
\pdfoutput=1 

\begin{document}
\title{Core collapse and horizontal-branch morphology in galactic globular clusters}
\titlerunning{Core collapse and HB morphology in GGCs}
\author{Mario Pasquato
\inst{1,2} \and Gabriella Raimondo \inst{3} \and Enzo Brocato \inst{4} \and Chul Chung \inst{1}
\and Anthony Moraghan \inst{1} \and Young-Wook Lee \inst{1}}
\institute {Department of Astronomy \& Center for Galaxy Evolution Research,
Yonsei University, Seoul 120-749, Republic of Korea\\
\and Yonsei University Observatory, Seoul 120-749, Republic of Korea\\
\and INAF-Osservatorio Astronomico di Teramo, Mentore Maggini s.n.c., 64100 Teramo, Italy \\
\and INAF-Osservatorio Astronomico di Roma, Via Frascati 33, 00040, Monte Porzio Catone, Italy}
\date{Received XXXX/ Received YYYY}

\abstract{Stellar collision rates in globular clusters (GCs)
do not appear to correlate with horizontal branch (HB) morphology, suggesting that dynamics does
not play a role in the second-parameter problem. However, core densities and collision rates derived
from surface-brightness may be significantly underestimated as the surface-brightness profile
of GCs is not necessarily a good indicator of the dynamical state of GC cores. Core-collapse
may go unnoticed if high central densities of dark remnants are present.}{We test whether GC
HB morphology data supports a dynamical contribution to the so-called second-parameter effect.}{{To remove first-parameter dependence }we fitted
the maximum effective temperature along the HB as a function of metallicity in a sample of $54$
Milky Way GCs. We plotted the residuals to the fit as a function of second-parameter candidates, {namely dynamical age and total luminosity}. We
considered dynamical age (i.e. the ratio between age and half-light relaxation time)
among possible second-parameters. We used {a set of direct N-body simulations, including ones} with dark remnants to illustrate how core density peaks, due to core collapse, in a dynamical-age range similar to
that in which blue HBs are overabundant with respect to the metallicity expectation, {especially for low-concentration initial conditions}.}{
GC total luminosity shows nonlinear behavior compatible with the
self-enrichment picture. However, the data are amenable to a different interpretation based on
a dynamical origin of the second-parameter effect. Enhanced mass-stripping in the late red-giant-branch phase
due to stellar interactions in collapsing cores is a viable candidate mechanism. In this picture,
GCs with HBs bluer than expected based on metallicity are those undergoing
core-collapse.}{}  \maketitle

\section{Introduction}
The color distribution of horizontal branch (HB) stars differs greatly among globular clusters
\citep[GCs; see, for example,][]{2002A&A...391..945P}. Since the 1950s, when metal abundances began to be determined for {GCs}, it was recognized that the first parameter responsible for this is metallicity \citep{1960ApJ...131..598S}. However, it was also realized that it alone cannot fully account for
the observed HB morphology, since observations showed that they can differ in the average position along the color coordinate in the HR diagram, in the overall extension, and in further additional features such as the presence of gaps. The task of singling out the mechanisms and the related observables that drive this diverse morphology, besides metallicity, is known as the second-parameter problem\footnote{Even though we retain this terminology for historical reasons, the implied assumption that a single second parameter can account for all the characteristics of the distribution of stars on the HB is unjustified in light of the current understanding of the issue.}.
Over the years several candidate quantities have been claimed to be either the second parameter or a second parameter \citep[see for a review][]{2009Ap&SS.320..261C, 2010A&A...517A..81G}. A (probably not exhaustive) list of these candidates are: age \citep[][]{1994ApJ...423..248L,
dotter_acs_2010}, helium abundance \citep[][]{2005ApJ...621L..57L, 2007A&A...474..105B,
2007ApJ...661L..53P, 2008ApJ...677.1080Y}, concentration \citep[][]{1993AJ....105.1145F}, binaries \citep[][]{2012arXiv1212.3063L}, and total GC mass \citep[][R$06$ in the
following]{recio-blanco_multivariate_2006}. More recently, this long-standing issue attracted renewed interest after the discovery of multiple stellar populations in GCs \citep[][among several others]{2004ApJ...605L.125B, 2005ApJ...621..777P, 2007A&A...474..105B,2007ApJ...661L..53P, 2008ApJ...677.1080Y}. However, a comprehensive picture is still lacking, and it is now clear that a single quantity is not capable of determining HB morphology in its whole complexity, so that more than one parameter is likely to play a role.

After the seminal work of \cite{1978ApJ...225..357S}, the second parameter problem acquired a greater significance as a probe of the formation history of the Galaxy.
\cite{1978ApJ...225..357S} pointed out that GCs with red HBs are usually found at galactocentric distances $R_{GC} > 8$kpc, while their occurrence is rare at smaller $R_{GC}$. This led to the well-known inner halo-outer halo dichotomy, the outer halo GC population being accreted over an extended period of time according to the merging paradigm for galaxy formation. In this context, a quantity capable of explaining the differences in the mean temperature of the HB between inner- and outer-halo GCs in the Milky Way (MW) is often called a global second parameter. Amongst the various candidates, age is currently regarded as the most appropriate global second parameter \citep[][]{1994ApJ...423..248L, dotter_acs_2010}.
%The latter authors measured the median color difference between HB and the RGB from ACS/HST photometry of a sample of GGC, and demonstrated that this color difference correlates with age.

A quantity is instead termed as the local second parameter if it is capable of explaining the extension of the HB within a given GC. The discovery of multiple main sequences in GCs, together with the presence of extremely blue HB stars, has given strength to helium abundance as a local second parameter. A self-enrichment in helium and multiple generations of stars within an individual cluster can explain features such as tails and multimodalities in the HB \citep[see][]{2002A&A...395...69D,2010A&A...517A..81G,2013ApJ...762...36J}.

In this picture, dynamical processes (e.g. stellar encounters, binarity, stellar rotation) would not be involved in the global second-parameter issue, but their importance might be paramount as local second parameters, i.e. in connection to the extension of the HB (range of temperatures) rather than to its overall position in the HR diagram. After a {GC} forms, the negative heat capacity of the cluster self-gravity draws the cluster toward core collapse, unless another energy source is present. The central region or core will collapse to very large densities within a few relaxation times while the outer regions expand, and stars evaporate from the system. The standard energy source that is invoked to retard and to stop a further collapse is binaries \citep[e.g.][]{1989Natur.339...40G}. The contraction of the core via two-body relaxation increases its density. The core density determines the interaction rate in the core, which affects the single and binary star interaction probabilities at a given time. {Core density is difficult to measure observationally because of the presence of invisible dark remnants. Moreover, the fraction of neutron stars and stellar-mass black holes that are expelled from GCs shortly after forming, due to natal-kick velocities, is poorly constrained.} Due to this complexity in the evolution of dense star clusters, the only way to understand more about this picture is to perform numerical simulations including the physical processes mentioned above with a large number of stars of different types, i.e. visible stars and remnants.

The HB temperature distribution in Galactic GCs (GGCs) is usually interpreted as evidence of a dispersion in HB masses
due to varying amounts of mass loss in the red-giant branch (RGB) phase. It is usually expected that the leading term in RGB mass-loss is due to stellar winds \citep[][]{1977A&A....57..395R} and to reproduce the HB, an RGB mass-loss distribution is assumed accordingly, usually a Gaussian as first proposed by \citet{1973ApJ...184..815R}, and later used by
\citet{1994ApJ...423..248L, 2000A&AS..146...91B, 2002ApJ...569..975R}. However, mass loss caused only by stellar wind is unable to explain the observed HB morphologies of some GCs in the MW \citep[e.g., the hottest blue HB stars in NGC 2808, see][]{2011MNRAS.410..694D}. This points to another mass-loss process that may well be dynamical in origin, given that the physics of mass-loss is still poorly understood \citep[e.g.][]{2009AJ....138.1485D}.
A numerical simulation study of RGB mass loss through encounters with white dwarfs and main sequence stars
was performed by \citet{1991ApJ...381..449D} who characterized the dependence of mass-loss on the impact parameter and relative
velocity of the encounter using a smoothed particle hydrodynamics (SPH) code. In this paper we report a finding that supports the view that mass-loss is related to the internal GC dynamics, in particular to core-collapse.
%We use a statistical approach based on Exploratory Data Analysis (EDA) and Non-Linear Dimensionality Reduction (NLDR) to uncover a non-linear relation between blue HBs and GC dynamical age, whose interpretation as a dynamical effect is supported by comparison with core density evolution as a function of dynamical age, as obtained using a direct N-body simulation.
We will investigate mass-loss through an encounter scenario in more detail in a future work (Moraghan et al., in preparation).

%Summarize the following?

\section{HB morphology and core density}
R$06$ used \emph{Hubble} Space Telescope photometry of a sample of $54$ Milky Way GCs to determine the most probable second-parameter candidate for HB morphology, as quantified by the maximum effective temperature along the HB. In the present paper we introduce a new parameter, the cluster dynamical age, defined as the ratio between cluster age and half-light relaxation time and use the same data of R$06$ together with direct N-body simulations to investigate a possible correlation between HB morphology and GGC core density.

In general, different ways to quantify HB morphology may probe different physical processes and be subject to different statistical uncertainties and biases. While maximum effective temperature is an optimal indicator in the multivariate sense of the study by \citet{Chatto}, being based on an extreme value it is intrinsically noisier than central tendency estimates such as the mean and median. It is however more suitable to probe the physical processes that extend the HB towards high temperatures, which are more likely to be of dynamical origin as we argued above and show in the following. {\cite{2012AJ....144..126D} found that maximum HB temperature from the R$06$ dataset correlates with ultraviolet colors obtained from GALEX integrated photometry, particularly with those involving far ultraviolet magnitudes. Therefore, even though maximum HB temperature based on optical colors, such as that of R$06$, might be underestimated in some cases \citep[see][for example with NGC $2808$]{2011MNRAS.410..694D}, this quantity is still a useful parameter for characterizing HB morphology.}

% as shown for example in the case of NGC $2808$ by \cite{2011MNRAS.410..694D}

In their paper R$06$ compute the one-to-one linear correlation of the maximum effective temperature along the HB with metallicity, total GC luminosity, age, relaxation time, central density, and several other variables, and use the Principal Component Analysis \citep[PCA; ][]{PCA} technique to check whether the best second-parameter may be a linear combination of a subset of these quantities. They find that the total luminosity is by far the most significant parameter in determining HB morphology after metallicity, and interpret this finding in terms of the GC self-pollution picture \citep[][]{2002A&A...395...69D}, where more massive (and therefore luminous) GCs managed to retain the helium-enriched ejecta of asymptotic giant branch stars. The authors conclude that {linear} correlations with central density, relaxation time, and the collisional parameter $\Gamma_{col}$ are marginally significant at best. This finding apparently rules out the role of stellar dynamics in the second-parameter problem.

{In the next Section, we show} that the cluster dynamical age affects HB morphology in a nonlinear way: linear techniques such as PCA do not perform well when nonlinear associations between parameters are present \citep[e.g. see][]{2010arXiv1001.1122G}, so this may be a reason why this effect was overlooked in R$06$, even though both parameters used to define dynamical age ({chronological} age and relaxation time) were considered separately.

\begin{figure}
\includegraphics[width=0.99\columnwidth,natwidth=800,natheight=500]{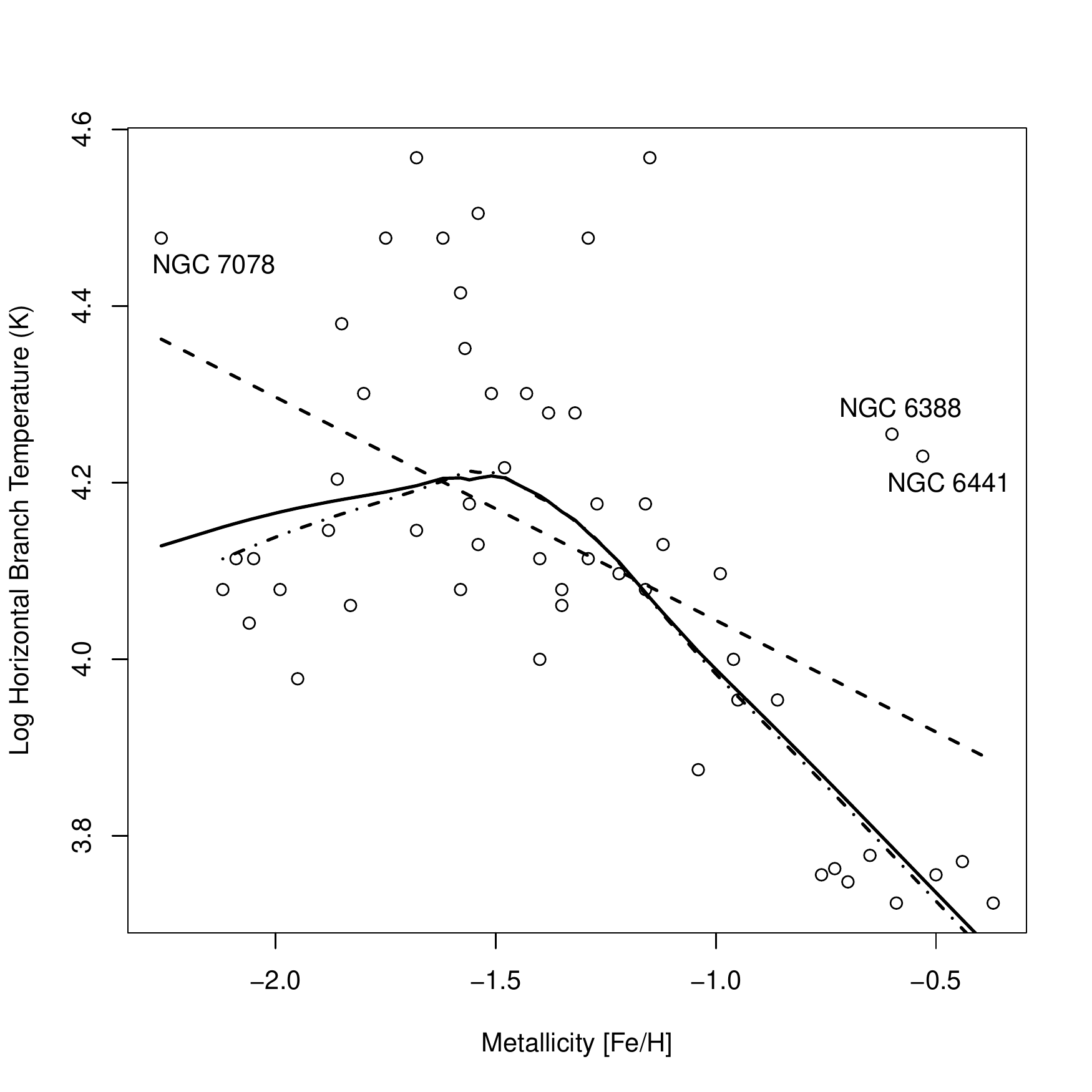} \caption{Relation
between maximum HB temperature and metallicity for R$06$ GCs. The superimposed {straight dashed line is a linear least-square fit, and the} solid line is a local polynomial
regression smoothing. Three points appear visually as outliers to the relation:  NGC $7078$, NGC $6388$, and NGC $6441$. The {dash-dotted} line is the fit obtained without including them in the data-set. The difference with respect to the solid line is relatively small, due to the robust algorithm used for local polynomial regression (see text for details).\label{f7}} \end{figure}

The relation between maximum effective HB temperature and metallicity for the R$06$
GC sample is shown in Fig.~\ref{f7}, {alongside an unweighted linear least-square fit (dashed straight line)}.
We {also} applied local polynomial smoothing to extract the
overall trend (solid {and dash-dotted lines} in Fig.~\ref{f7}) using the lowess() routine in the R programming language.
This is a nonparametric technique where a polynomial is fit to the data
inside a moving window, and makes use of an iterative re-weighing scheme to lower the influence of outliers, while recovering the trend from the data without the need for model assumptions.
The dependence on metallicity is stronger for high-metallicity GCs.
The slope decreases (and {is even} inverted) for low-metallicity GCs, while the scatter increases. {However, as shown by the straight linear regression line, the overall slope is negative.}

\begin{figure} 
\includegraphics[width=0.99\columnwidth,natwidth=800,natheight=500]{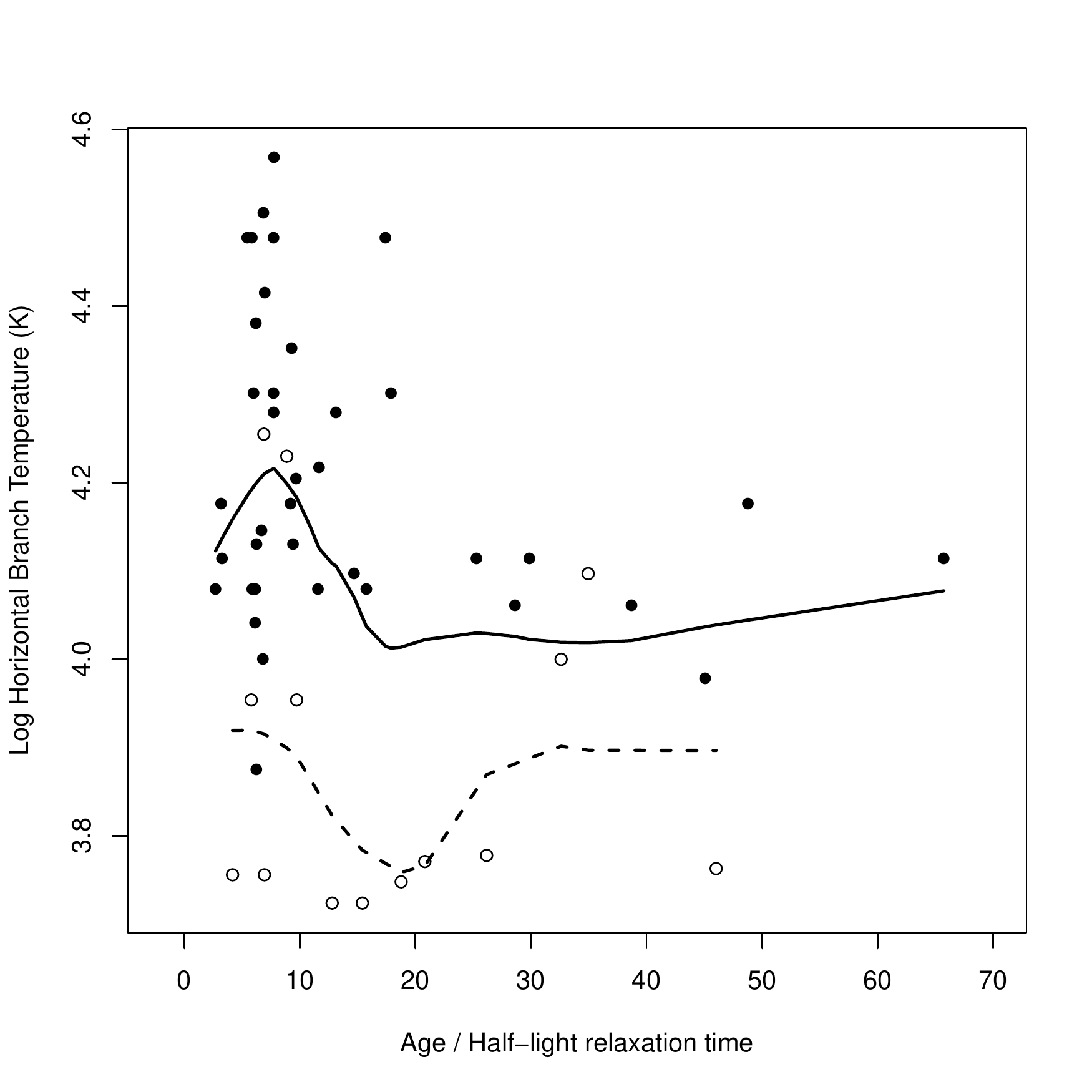} \caption{Logarithm of maximum effective HB temperature as a function of dynamical age for the R$06$ sample, with superimposed local polynomial regression line (solid) showing the overall trend. High-metallicity GCs (the sample's top $25\%$ in metallicity, {i.e. the GCs with $[Fe/H] \geq -1.0$}) are represented by empty circles, the rest by filled circles. Temperature peaks in the $5$-$10$ relaxation-time range, but high metallicity GCs do not follow the trend. Their trend with dynamical age is shown by the dashed line. They are generally lower in HB temperature, because the main parameter affecting their HB temperature is metallicity, not dynamical age.  \label{logTonly}} \end{figure}

\begin{figure} 
\includegraphics[width=0.99\columnwidth,natwidth=800,natheight=500]{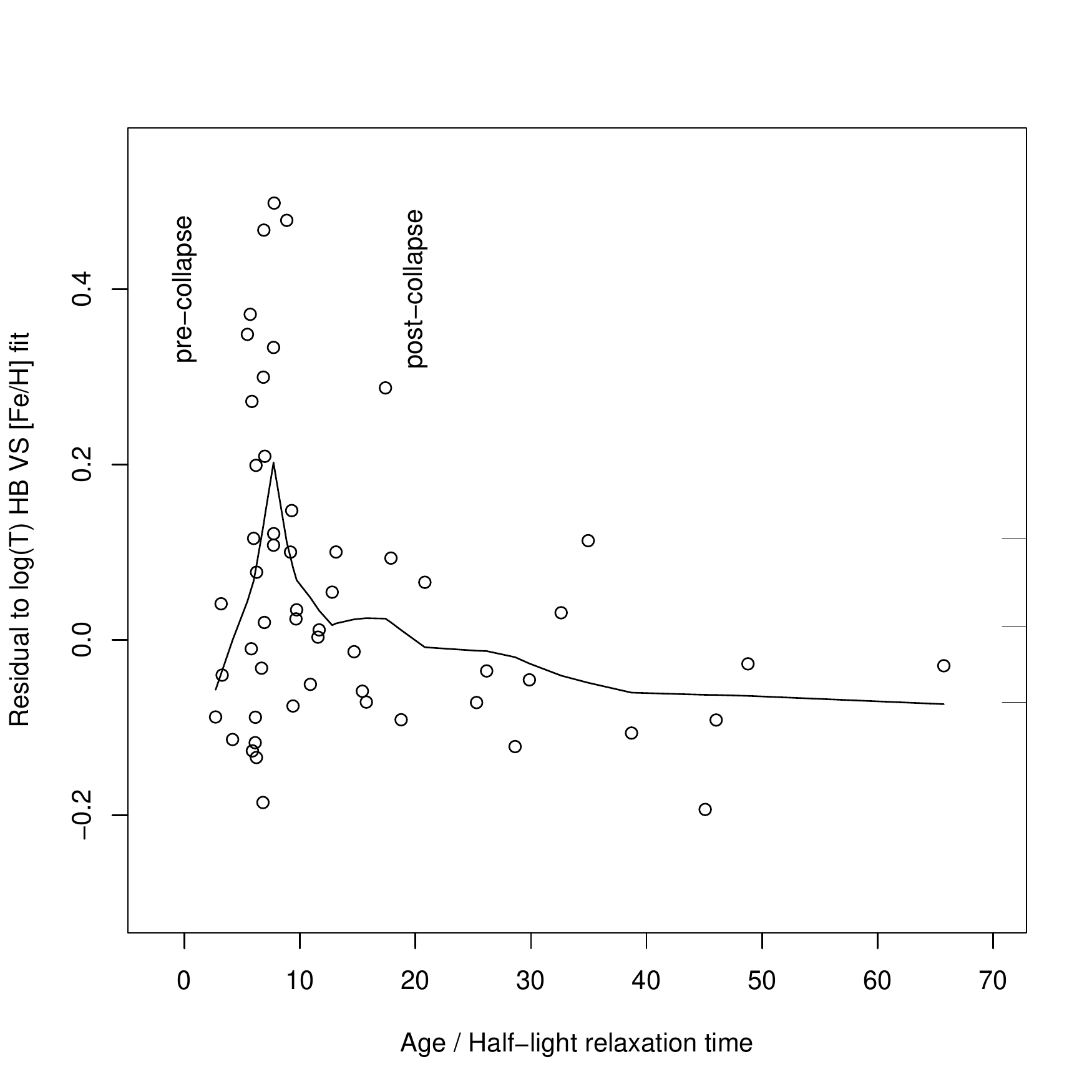} \caption{Residuals to the
local polynomial regression plotted in Fig.~\ref{f7} {(solid line in that figure)} as a function of GC age to relaxation-time
ratio. Core-collapse is {found}
to occur for a ratio of about $5$ to $10$ depending on the mass spectrum and {initial concentration (see simulations in the following)}. The superimposed local polynomial regression line (dashed) shows a peak in this
age range. For comparison, the ticks on the vertical axis at the right side of the plot show (from bottom to top) the first quartile, the median, and the third quartile of the distribution of residuals.\label{f9}} \end{figure}

\begin{figure} 
\includegraphics[width=0.99\columnwidth,natwidth=800,natheight=500]{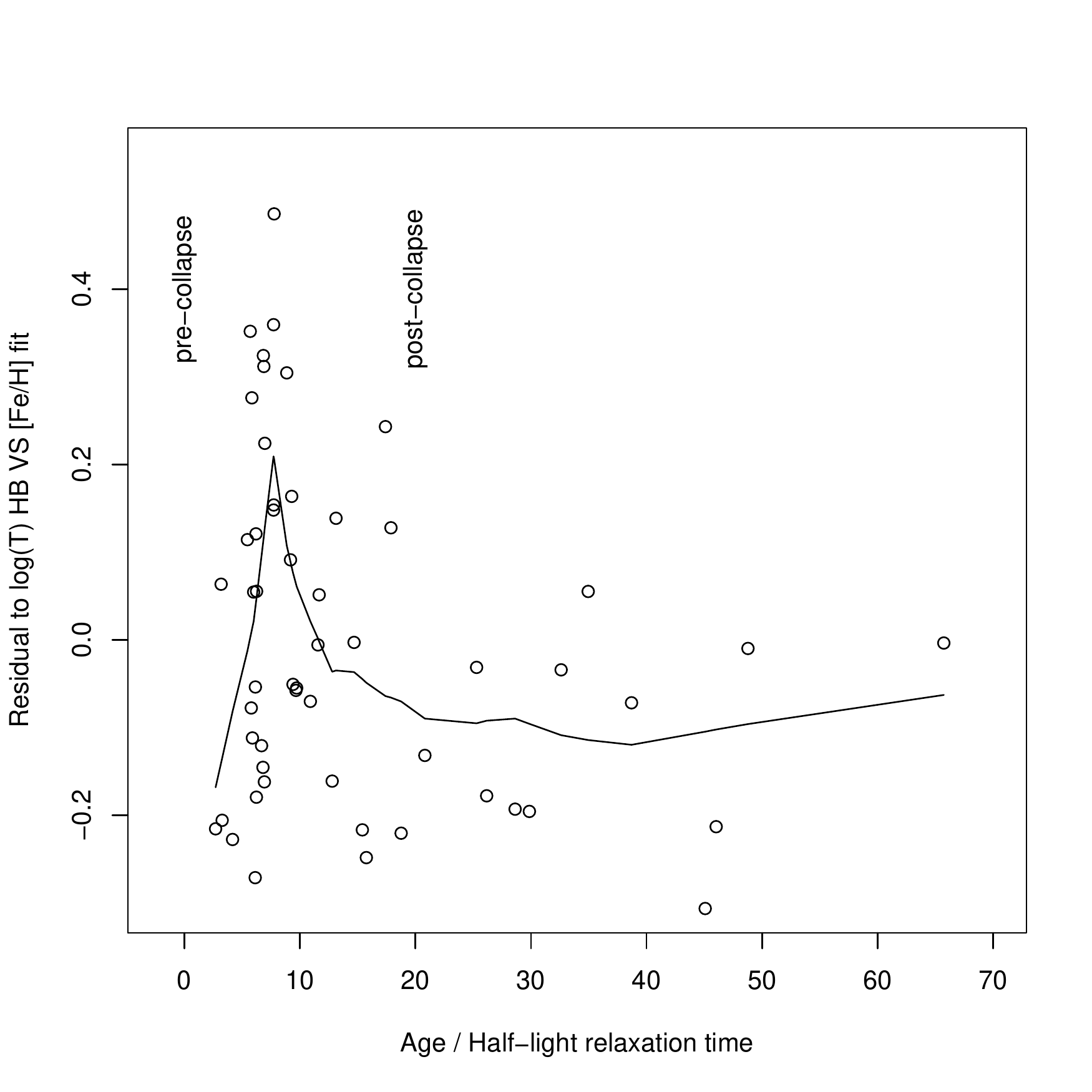} \caption{Residuals to the
linear regression plotted in Fig.~\ref{f7} as a function of GC age to relaxation-time
ratio (as in Fig.~\ref{f9}).  The superimposed local polynomial regression line (dashed) is also obtained as in Fig.~\ref{f9} and shows an overall very similar behavior of the residuals. \label{linearres}} \end{figure}

\subsection{Dependence on dynamical age}
{We define dynamical age as the ratio between chronological age and half-light relaxation time taken from \cite{1996AJ....112.1487H}. Absolute age is obtained from relative ages \citep[][]{2009ApJ...694.1498M} normalized to $13$ Gyr, where $5$ clusters lacking a relative age measurement have been mean-substituted. In this section we quantify the effect of this parameter on maximum HB temperature.}

{In their analysis, R$06$ find that metallicity is the most influential parameter driving maximum HB temperature, based on linear measures of association (i.e. the correlation coefficient). In the following, we will therefore remove its effect by subtracting the temperature trend with metallicity before looking for a second parameter. It is however instructive to look directly at the relation between maximum HB temperature and dynamical age, uncorrected for metallicity, as shown in Fig.~\ref{logTonly}. It is clear that a global trend is present (as shown by the smooth solid line in Fig.~\ref{logTonly}), with an increase of the maximum HB temperature in the dynamical-age range of $5$-$10$. However, the GCs in the highest-metallicity $25\%$ of the sample ({that also happen to have $[Fe/H] \geq -1.0$}; empty circles) follow a completely different trend (dashed line) and have on average lower HB temperatures. This is expected based on Fig.~\ref{f7}, as high-metallicity GCs show a very strong linear dependence of maximum HB temperature on metallicity. In the following we will therefore compare candidate second-parameters to the residuals to the temperature-metallicity relation instead of to temperature directly.}

Figure \ref{f9} shows a plot of the residuals to
the $\log T$ VS $[Fe/H]$ relation {(obtained based on a local polynomial fit, as explained above)} as a function of dynamical age. It turns out that clusters whose age is about $5$ to
$10$ relaxation times have large residuals (i.e. bluer HBs than expected based on metallicity),
while both dynamically-younger and older GCs have smaller residuals. {A similar result is obtained if residuals to the linear fit to the $\log T$ VS $[Fe/H]$ relation are considered instead, as shown in Fig.~\ref{linearres}. This is due to the fact that residuals to the linear fit and residuals to the polynomial fit are strongly correlated, with a linear correlation coefficient of $0.90$.}

%Moreover, also a direct comparison of the maximum HB temperature with dynamical age shows an overabundance of hot (blue) horizontal branches in the $5$-$10$ relaxation times age range, as shown in Fig.~\ref{logTonly}. The effect is less visible, though, because the metallicity dependence is not subtracted (i.e. the \emph{second parameter} effect is less visible if dependence on the \emph{first parameter}, metallicity, is not removed). In particular, HB temperatures of high-metallicity clusters are driven mainly by metallicity, so that they show no trend with dynamical age, as shown by the light-gray line in Fig.~\ref{logTonly}.

Core density is expected
to peak {in the $5$-$10$ relaxation times age range}, due to core collapse {(see discussion below)}. Several studies \citep[e.g. see][]{1997AJ....113..706B, 2001AJ....121..916T} suggest that core density is a good candidate
for the second-parameter role, but when the residuals are compared directly to the central density,
R$06$ find no strong correlation.  This may be due to the fact that central density estimates,
such as those found in \cite{1996AJ....112.1487H}, are based on surface brightness profiles that
are not good at tracing mass, especially in the central regions where heavy dark remnants such
as neutron stars and stellar-mass black holes tend to segregate \citep[][]{2010ApJ...708.1598T}.
The same applies to expected rates of stellar collisions estimated from central velocity dispersions
and central surface brightness, i.e. $\Gamma_{col}$.

{We run a set of direct N-body simulations using the Nbody $6$ code \citep[][]{2010gnbs.book.....A} in order to obtain the evolution of core density as a function of dynamical age. We then compare it with that of the residuals to the $\log T$-$[Fe/H]$ fit. Table~\ref{simu} shows a summary of our simulations. They all contain a number of particles in the range $16$ to $32$ thousand, start from mass-unsegregated King-model initial conditions, and probe a range of discrete mass-spectra. {Our models are isolated, i.e. no tidal forces from the galaxy have been included}. The time at which core-collapse occurs, as well as its depth, is sensitive on the ingredients included in the simulation, such as the initial concentration and especially the binary fraction. However, the goal of the present paper is to show the presence of a pattern in GC data that can be interpreted in terms of core-collapse influencing HB morphology, so we do not probe the full simulation parameter space. In particular, our simulations start from low initial concentrations and have a primordial binary fraction of $0$\footnote{However, binaries are formed dynamically at core collapse.}. While this is unrealistic, the goal of the simulations we present here is limited to illustrate how the temporal evolution of core density in GCs is remarkably similar to that of HB temperature (after removing metallicity dependence) without the need of a fine tuning of initial conditions.}

{In our analysis we define the core radius as the three-dimensional radius at which density drops by a factor $2$ with respect to the central density. We then calculate the average mass-density within it. The simulations contain either three or four classes of stellar masses, and throughout our set we explore different number ratios between the classes. In all simulations the lightest class is the most abundant in number, in order to represent visible stars (luminous main sequence, RGB and HB). The heavier classes can be understood as invisible remnants, such as neutron stars and stellar-mass black holes, and different number ratios between the classes represent different retention fractions of dark remnants. Dynamical age is computed, for each simulation snapshot, by dividing its chronological age (in N-body units) by the half-mass relaxation time.
{We have adopted two different prescriptions for calculating the half-mass relaxation time: we either take the initial value of the half-mass relaxation time as obtained by NBODY$6$ based on the initial conditions, or we recalculate it for each snapshot based on the actual half-mass radius and total mass of the model, according to the definition by \cite{1987gady.book.....B}. In the following we will refer to prescription (a) and (b) respectively. Since our models expand indefinitely (due to the lack of tidal limitation), half-mass radii tend to grow over time, producing an increasing half-mass relaxation time. In prescription (b) we first calculate the three-dimensional half mass radius $r_{h_{all}}$ of all the particles, and then consider only those contained within three times $r_{h_{all}}$. While calculating the relaxation time self-consistently as in prescription (b) is more physically ground, there are, however, great uncertainties about relaxation times obtained from the observational data and on how to measure dynamical age in a consistent way both in simulations and observations \citep[][]{1993ASPC...50..373D, 1996AJ....112.1487H, mclaughlin_resolved_2005}. Therefore we consider both prescriptions (a) and (b), and find that the differences are not dramatic, as we show in the following.} 
Figure \ref{simulazi} compares the temporal evolution of core mass-density in direct N-body simulations with the residuals plotted in Fig.~\ref{f9}. 

\begin{table}
\caption{Summary of simulation runs in this paper. The simulations either have three or four classes of mass. The number ratios and the mass ratios are given in the relevant columns in the table. The initial conditions are a King-model with dimensionless potential $W_0$ as listed in the table. Primordial binary fraction is consistently $0$ over the whole set of simulations.}
\label{simu}     
\centering                                      
\begin{tabular}{c c c c c}          
\hline\hline                       
ID & Stars & $W_0$ & Mass ratios & $N$ ratios \\ 
\hline                                  
    32k3cW02 & $32 \cdot {10}^3$ & $2$ & $1:2:3$ & $99:0.5:0.5$\\      % inserting body of the table
    16k3cW02 & $16 \cdot {10}^3$ & $2$ & $1:2:3$ & $98:1:1$\\
    16k4cW02  & $16 \cdot {10}^3$ & $2$ & $1:2:3:4$ & $97.5:1:1:0.5$\\
    16k4cW02 & $16 \cdot {10}^3$ & $2$ & $1:2:3:4$ & $97.75:1:1:0.25$\\
    16k3cW04 & $16 \cdot {10}^3$ & $4$ & $1:2:3$ & $98:1:1$ \\
    16k4cW04 & $16 \cdot {10}^3$ & $4$ & $1:2:3:4$ & $97:1:1:1$ \\
\hline                                             %inserts single line
\end{tabular}
\end{table}

%  and corresponding number fraction $99:0.5:0.5$,
%so that the lightest class represents visible stars, and the heavier classes represent neutron
%stars and/or stellar-mass black holes. The mass ratios were adjusted so that if the light
%component is slightly less than the typical turn-off mass of a GC (i.e. $0.8$ $M_\odot$), then the
%heavier ones are set to exceed the Chandrasekhar mass (i.e. about $1.44$ $M_\odot$).
%The initial conditions are a King model with low concentration ($W_0 = 2$) and no primordial binaries.
%We make use of the three-dimensional radius containing $10\%$ of the total number of particles to define a sphere within which we compute the average density. This choice is justified because we want to limit the fluctuations associated to a definition of the core radius based on half-peak density or King-model fitting.
{Core density obtained from the simulations is shown as a function of dynamical age, i.e. the ratio between age and half-light relaxation time in the bottom panel of
Fig.~\ref{simulazi} {(left, prescription (a); right, prescription (b))}, while the top panel shows the residuals to the $\log T$ VS $[Fe/H]$ relation as in Fig.~\ref{f9}.
The gray shaded area is the envelope of all the simulations in Tab~\ref{simu}, and the black solid line is the global average.
Core density, in all simulations in our set, peaks around $5$-$10$ initial half-light relaxation times, when the self-similar collapse is halted by the formation of (at least) a binary due to three-body interactions.} This is a general pattern that emerges in simulations of GCs irrespective of details related to the mass-function, so our choice of the mass-classes and number ratios discussed above does not change the qualitative evolution of density, even though a different mass spectrum {will} yield a quantitatively different evolution. {However}, it is clear from Fig.~\ref{simulazi} that
there is at least a qualitative similarity between the evolution of core density as a function
of dynamical age and the residuals to the $\log T$ VS $[Fe/H]$ relation, and any further investigation of the details of the core-density evolution is beyond the scope of this paper.

\begin{figure} 
\includegraphics[width=0.99\columnwidth,natwidth=800,natheight=500]{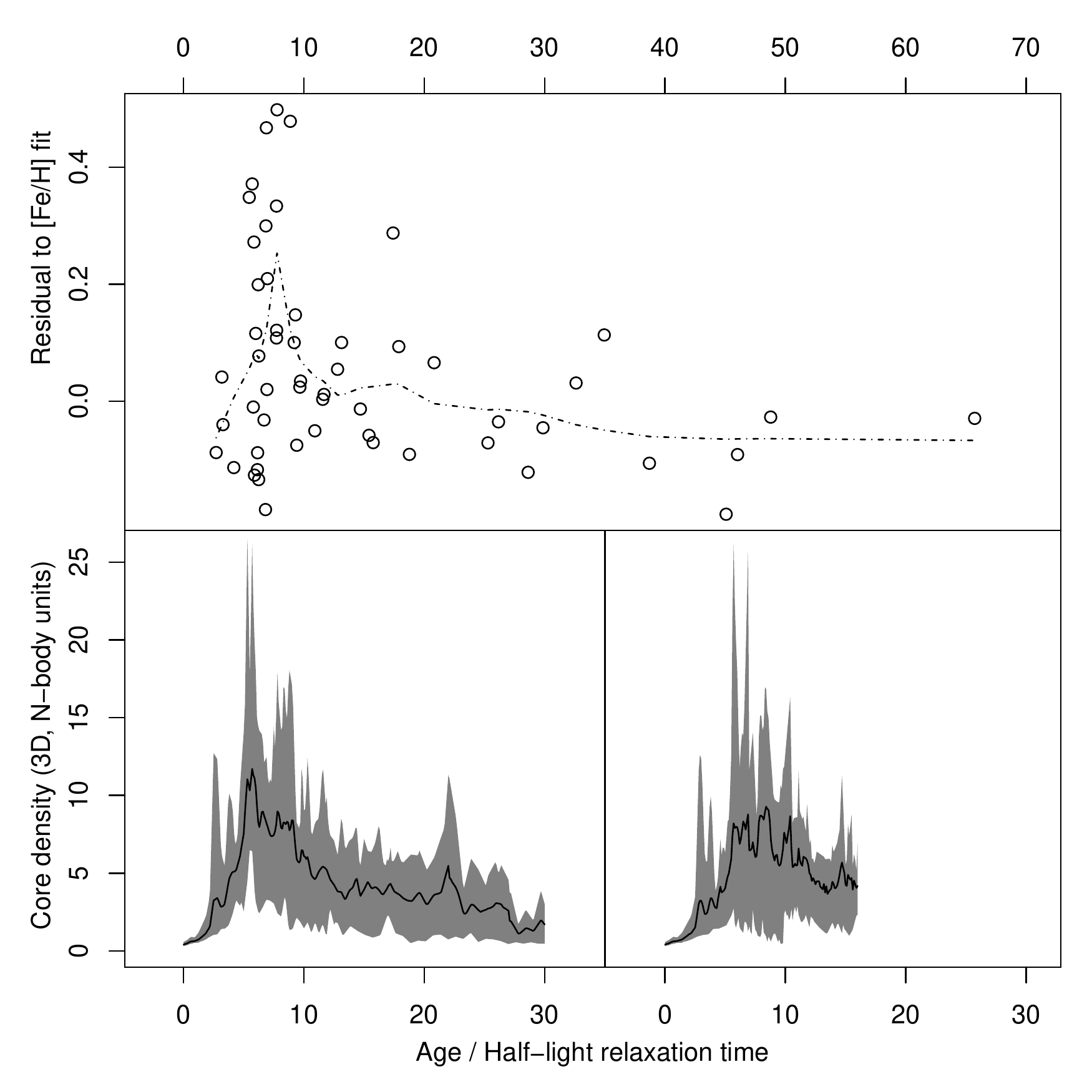} \caption{{Top panel:} Residuals to the metallicity fit associated to each GC, as in
Fig.~\ref{f9}. {Bottom panel, left:} Core
mass-density in {our set of direct N-body simulations of GCs (see Tab.~\ref{simu})} as a function of {dynamical age}. The gray shaded area is the envelope encompassing all simulations. The black solid line is the global average. {The dynamical age was computed using the initial half-mass relaxation time (prescription (a) in the text).} {Bottom panel, right: same as left panel, but half-mass relaxation time was recomputed for each snapshot based on \cite{1987gady.book.....B} definition (prescription (b) in the text). In both cases }
density peaks between $5$ and $10$ initial half-mass relaxation times. {Upper and lower panels all share the same horizontal axis, but in the upper panel the abscissa values extend more, so we put axis labels on its upper side for improved readability}. \label{simulazi}} \end{figure}

\subsection{Dependence on luminosity}

\begin{figure}
\includegraphics[width=0.99\columnwidth,natwidth=800,natheight=500]{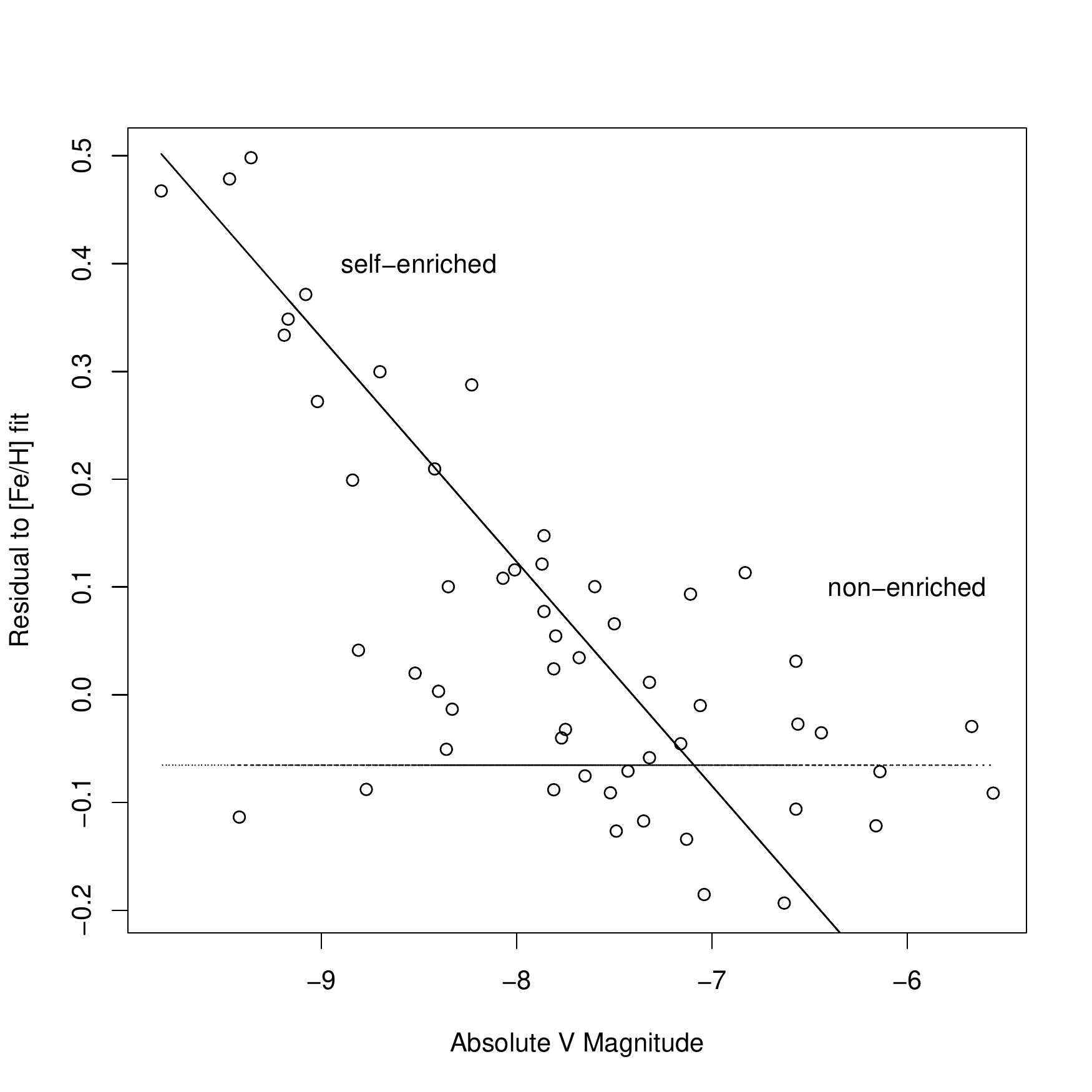} \caption{Residuals
to the local polynomial regression plotted in Fig.~\ref{f7} as a function of GC absolute
magnitude. Highly luminous clusters have larger residuals that increase with luminosity,
while clusters less luminous than about $-8$ $M_V$ display no correlation between scatter and
luminosity. If the current luminosity of a GC correlates with its initial mass, this pattern
can be interpreted as an effect of self-enrichment, that was effective only for clusters above
a given mass threshold.\label{f8}} \end{figure}

Figure \ref{f8} shows the residuals to the $\log T$
VS $[Fe/H]$ relationship (as obtained by local polynomial regression, see Fig.~\ref{f7})
as a function of the absolute magnitude $M_V$. At about $M_V = -8$, a threshold exists, and
different behaviors are displayed by GCs above and below it. Low-luminosity clusters do
not display any relation between luminosity and the residuals to the $\log T$ VS $[Fe/H]$ relationship,
while high-luminosity ones do. {A possible interpretation is based on the self-enrichment picture,
where more massive clusters were able to retain more stellar ejecta, thus self-enriching more efficiently.
However, current luminosity is not directly related to the original mass the cluster had at formation, as both the current mass-to-light ratio and the amount of mass lost during cluster lifetimes is uncertain \citep[see][and references therein]{2009ApJ...698L.158K}. It is anyway tempting to speculate that the magnitude threshold we observe in Fig.~\ref{f8} corresponds to the minimum initial mass needed for self-enrichment
to be efficient}. Models of GC formation such as \citet{recchi_self-enrichment_2005} predict a
typical mass threshold for self enrichment that would correspond to {an absolute magnitude $M_V$} of about $-7.8$,
compatible with our analysis. Masses above the threshold would correspond to higher
escape velocities, causing a better retention of chemically enriched gas. On the other hand,
for any value of the mass below the threshold, enrichment would not occur, explaining why mass
is no longer an influential variable and why the scatter becomes independent of magnitude.

{Our analysis of the data is thus also compatible with the self-enrichment hypothesis, but
as we have shown above this does not rule out a dynamical interpretation of the second-parameter
effect. In other words, the data seem compatible with both scenarios.}

\section{Comparison with other indicators of the GC dynamical state}
The previous discussion can be summarized by stating that core collapse is a second-parameter of HB morphology, in the sense that GCs with bluer HBs (with respect to the metallicity prediction) are likely to be undergoing core-collapse. However, in Fig.~\ref{densrho} we show that there need not be visible signs of core collapse in the surface-brightness profile of a GC when it is in the dynamical age range of $5$ to $10$ half-light relaxation times. We plot the logarithmic core density of the R$06$ GCs from \citet{1996AJ....112.1487H} as a function of dynamical age. {The pattern that emerges shows a very mild peak at a dynamical age of about $15$-$20$, as opposed to the stronger peak at $5$-$10$ in Fig.~\ref{f9}. The density of visible stars probably correlates with the overall density, but with a large scatter. Sources of scatter are the shot noise from a limited number of bright stars, and differences in the present-day mass function and remnant abundance in different GCs, all contributing into making the surface-brightness profile a poor indicator of the dynamical status of the core.}

Core-collapse is also often expected to produce a cusp in the surface-brightness profile of GCs, to the point that cuspy profiles are sometimes named collapsed-core GCs. There is a growing body of evidence, however, that the presence of a cusp in the surface brightness profile is neither sufficient nor necessary to conclude that the core is undergoing gravitational collapse.  \citet{2010ApJ...708.1598T} use state-of-the-art direct N-body simulations to show that visible cusps may not form in GCs, resulting in light profiles that are well fit by a King model even during the core-collapse stage, if a realistic fraction of dark remnants (neutron stars and stellar-mass black holes) are included in the simulation. On the other hand, even if present, cusps are hard to detect without high angular resolution. In Fig.~\ref{densrho} filled circles represent clusters whose profiles are centrally cuspy. They are assigned a conventional value of the concentration parameter $c = 2.5$ in the \citet{1996AJ....112.1487H} catalog. Except for having an overall higher value of core density, cuspy GCs do not have a clear pattern in the picture, again showing that the presence of a cusp is a poor proxy for dynamical age.

\begin{figure}
\includegraphics[width=0.99\columnwidth,natwidth=800,natheight=500]{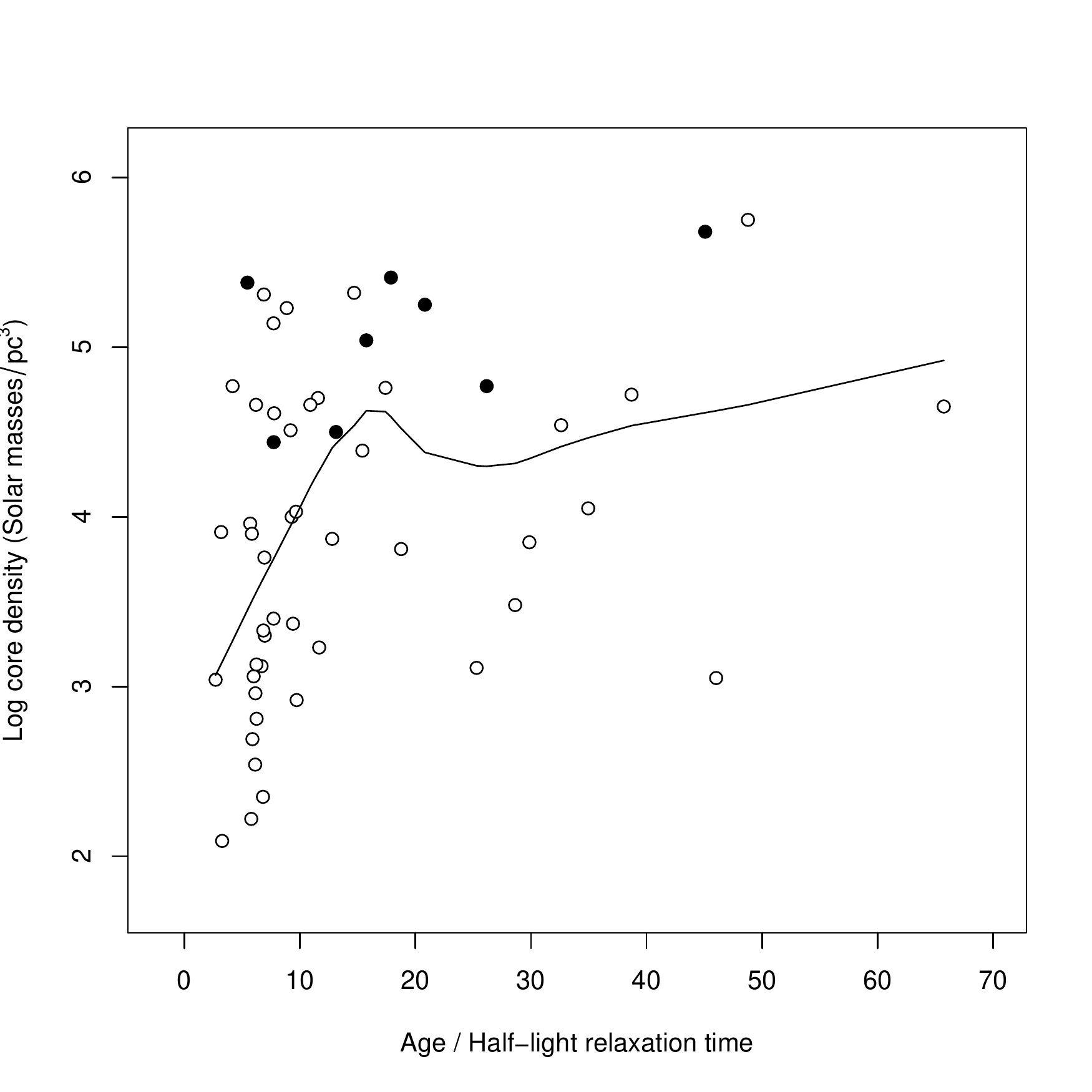} \caption{Log core density of the R$06$ GCs (estimated from central surface brightness) as a function of dynamical age. Filled circles are high-concentration GCs which have been assigned $c = 2.5$ in \citet{1996AJ....112.1487H}. The surface-brightness profiles of these GCs show a central cusp, and they are often termed collapsed-core even though the link between cuspy profiles and dynamically collapsed cores is uncertain. Density estimated from surface brightness {does actually show a mild peak at a dynamical age of about $15$-$20$ (solid smoothing line), but does not follow a pattern as distinctive as that of the overall density obtained from simulations}, and {the presence of a central cusp} is not a guarantee of old dynamical age.\label{densrho}} \end{figure}

\section{Conclusions and future prospects} 

We used data from R$06$ to show that internal dynamical processes can play a role in modeling the HB morphology in GGCs. In particular, GC dynamical age ({chronological age} measured in units of half-light relaxation time) affects HB morphology in a nonlinear way: GCs aged between $5$ and $10$ relaxation times have
bluer HBs than expected based on the metallicity prediction alone, while dynamically younger
and older GCs do not show large deviations. Since core-collapse is expected to take place in
{a similar} range of dynamical ages, we suggest the presence of a link between the increased central
density of collapsing cores and increased mass-loss in the RGB phase. However, core-collapse is
not necessarily apparent from the GC surface-brightness profile, so the concentration parameter
and/or the central surface-brightness are generally not good tracers of the GC dynamical state
and will only marginally correlate with blue HBs.

Given that the R$06$ sample is quite representative of the GGC population, ranging from metallicities of $-2.26$ to $-0.37$ and absolute magnitudes from $-5.56$ to $-9.82$, we may infer from it that a relation between dynamical age and exceedingly blue HBs is present in the whole {Galactic} population, and possibly in extragalactic GCs too. By looking at Fig.~\ref{f9} we see that about $20\%$ of the R$06$ GCs have ages in the $5$ to $10$ half-light relaxation time range and exceedingly blue HBs (i.e. residuals to the metallicity fit in the top quartile), so over a population of about $200$ GCs in the MW, we expect at least $40$ where core-collapse has a visible effect on HB morphology.

How exactly core-collapse affects RGB mass-loss is still an open question. For example,
\citet{2012arXiv1212.3063L} propose the presence of a companion (i.e. the HB star being,
or having been, part of a binary during its RGB evolutionary phase) as a candidate second-parameter.
The companion would enhance mass-loss from the primary by tidal interaction, making
the resulting HB star bluer. This theory is interesting in relation to our finding, as the
number and binding energy of binaries in a GC undergoing core collapse is expected to increase with time
\citep[][]{1987degc.book.....S}. Core-collapse would then indirectly affect HB morphology by
either creating binaries or reducing the separations of existing ones. Another possibility is that mass-stripping in the RGB phase is directly enhanced by stellar encounters, {that are more frequent} in a dense environment. We are currently modeling this process quantitatively using SPH simulations in order to provide a prediction of the mass-loss law at work in the high density environment of the core collapse GGCs.

\section*{Acknowledgements}
Support for this work was provided by
the National Research Foundation of Korea to the Center for Galaxy Evolution Research, and also
by the KASI-Yonsei Joint Research Program for the Frontiers of Astronomy and Space Science and the DRC program of Korea Research Council of Fundamental Science and Technology (FY 2012).
This work received partial financial support by INAF$-$PRIN$/$2010 (PI G. Clementini) and  INAF$-$PRIN$/$2011 (PI M. Marconi). {We wish to thank the referee for comments and questions that helped us clarify several important points.}

\bibliography{manuscript}

\begin{thebibliography}{42}
\expandafter\ifx\csname natexlab\endcsname\relax\def\natexlab#1{#1}\fi

\bibitem[{{Aarseth}(2010)}]{2010gnbs.book.....A}
{Aarseth}, S.~J. 2010, {Gravitational N-Body Simulations} (Cambridge University
  Press, Cambridge, UK)

\bibitem[{{Babu} {et~al.}(2009){Babu}, {Chattopadhyay}, {Chattopadhyay}, \&
  {Mondal}}]{Chatto}
{Babu}, G.~J., {Chattopadhyay}, T., {Chattopadhyay}, A.~K., \& {Mondal}, S.
  2009, \apj, 700, 1768

\bibitem[{{Bedin} {et~al.}(2004){Bedin}, {Piotto}, {Anderson}, {Cassisi},
  {King}, {Momany}, \& {Carraro}}]{2004ApJ...605L.125B}
{Bedin}, L.~R., {Piotto}, G., {Anderson}, J., {et~al.} 2004, \apjl, 605, L125

\bibitem[{{Binney} \& {Tremaine}(1987)}]{1987gady.book.....B}
{Binney}, J. \& {Tremaine}, S. 1987, {Galactic dynamics} (Princeton University
  Press, Princeton, NJ)

\bibitem[{{Brocato} {et~al.}(2000){Brocato}, {Castellani}, {Poli}, \&
  {Raimondo}}]{2000A&AS..146...91B}
{Brocato}, E., {Castellani}, V., {Poli}, F.~M., \& {Raimondo}, G. 2000, \aaps,
  146, 91

\bibitem[{{Buonanno} {et~al.}(1997){Buonanno}, {Corsi}, {Bellazzini},
  {Ferraro}, \& {Pecci}}]{1997AJ....113..706B}
{Buonanno}, R., {Corsi}, C., {Bellazzini}, M., {Ferraro}, F.~R., \& {Pecci},
  F.~F. 1997, \aj, 113, 706

\bibitem[{{Busso} {et~al.}(2007){Busso}, {Cassisi}, {Piotto}, {Castellani},
  {Romaniello}, {Catelan}, {Djorgovski}, {Recio Blanco}, {Renzini}, {Rich},
  {Sweigart}, \& {Zoccali}}]{2007A&A...474..105B}
{Busso}, G., {Cassisi}, S., {Piotto}, G., {et~al.} 2007, \aap, 474, 105

\bibitem[{{Catelan}(2009)}]{2009Ap&SS.320..261C}
{Catelan}, M. 2009, \apss, 320, 261

\bibitem[{{Dalessandro} {et~al.}(2011){Dalessandro}, {Salaris}, {Ferraro},
  {Cassisi}, {Lanzoni}, {Rood}, {Fusi Pecci}, \& {Sabbi}}]{2011MNRAS.410..694D}
{Dalessandro}, E., {Salaris}, M., {Ferraro}, F.~R., {et~al.} 2011, \mnras, 410,
  694

\bibitem[{{Dalessandro} {et~al.}(2012){Dalessandro}, {Schiavon}, {Rood},
  {Ferraro}, {Sohn}, {Lanzoni}, \& {O'Connell}}]{2012AJ....144..126D}
{Dalessandro}, E., {Schiavon}, R.~P., {Rood}, R.~T., {et~al.} 2012, \aj, 144,
  126

\bibitem[{{D'Antona} {et~al.}(2002){D'Antona}, {Caloi}, {Montalb{\'a}n},
  {Ventura}, \& {Gratton}}]{2002A&A...395...69D}
{D'Antona}, F., {Caloi}, V., {Montalb{\'a}n}, J., {Ventura}, P., \& {Gratton},
  R. 2002, \aap, 395, 69

\bibitem[{{Davies} {et~al.}(1991){Davies}, {Benz}, \&
  {Hills}}]{1991ApJ...381..449D}
{Davies}, M.~B., {Benz}, W., \& {Hills}, J.~G. 1991, \apj, 381, 449

\bibitem[{{Djorgovski}(1993)}]{1993ASPC...50..373D}
{Djorgovski}, S. 1993, in Astronomical Society of the Pacific Conference
  Series, Vol.~50, Structure and Dynamics of Globular Clusters, ed. S.~G.
  {Djorgovski} \& G.~{Meylan}, 373

\bibitem[{Dotter {et~al.}(2010)Dotter, Sarajedini, Anderson, Aparicio, Bedin,
  Chaboyer, Majewski, Mar{\'\i}n-Franch, Milone, Paust, Piotto, Reid,
  Rosenberg, \& Siegel}]{dotter_acs_2010}
Dotter, A., Sarajedini, A., Anderson, J., {et~al.} 2010, \apj, 708, 698

\bibitem[{{Dupree} {et~al.}(2009){Dupree}, {Smith}, \&
  {Strader}}]{2009AJ....138.1485D}
{Dupree}, A.~K., {Smith}, G.~H., \& {Strader}, J. 2009, \aj, 138, 1485

\bibitem[{{Fusi Pecci} {et~al.}(1993){Fusi Pecci}, {Ferraro}, {Bellazzini},
  {Djorgovski}, {Piotto}, \& {Buonanno}}]{1993AJ....105.1145F}
{Fusi Pecci}, F., {Ferraro}, F.~R., {Bellazzini}, M., {et~al.} 1993, \aj, 105,
  1145

\bibitem[{{Goodman} \& {Hut}(1989)}]{1989Natur.339...40G}
{Goodman}, J. \& {Hut}, P. 1989, \nat, 339, 40

\bibitem[{{Gorban} \& {Zinovyev}(2010)}]{2010arXiv1001.1122G}
{Gorban}, A.~N. \& {Zinovyev}, A. 2010, ArXiv e-prints 1001.1122

\bibitem[{{Gratton} {et~al.}(2010){Gratton}, {Carretta}, {Bragaglia},
  {Lucatello}, \& {D'Orazi}}]{2010A&A...517A..81G}
{Gratton}, R.~G., {Carretta}, E., {Bragaglia}, A., {Lucatello}, S., \&
  {D'Orazi}, V. 2010, \aap, 517, A81

\bibitem[{{Harris}(1996)}]{1996AJ....112.1487H}
{Harris}, W.~E. 1996, \aj, 112, 1487

\bibitem[{{Joo} \& {Lee}(2013)}]{2013ApJ...762...36J}
{Joo}, S.-J. \& {Lee}, Y.-W. 2013, \apj, 762, 36

\bibitem[{{Kruijssen} \& {Portegies Zwart}(2009)}]{2009ApJ...698L.158K}
{Kruijssen}, J.~M.~D. \& {Portegies Zwart}, S.~F. 2009, \apjl, 698, L158

\bibitem[{{Lee} {et~al.}(1994){Lee}, {Demarque}, \&
  {Zinn}}]{1994ApJ...423..248L}
{Lee}, Y.-W., {Demarque}, P., \& {Zinn}, R. 1994, \apj, 423, 248

\bibitem[{{Lee} {et~al.}(2005){Lee}, {Joo}, {Han}, {Chung}, {Ree}, {Sohn},
  {Kim}, {Yoon}, {Yi}, \& {Demarque}}]{2005ApJ...621L..57L}
{Lee}, Y.-W., {Joo}, S.-J., {Han}, S.-I., {et~al.} 2005, \apjl, 621, L57

\bibitem[{{Lei} {et~al.}(2012){Lei}, {Chen}, {Zhang}, \&
  {Han}}]{2012arXiv1212.3063L}
{Lei}, Z.-X., {Chen}, X.-F., {Zhang}, F.-H., \& {Han}, Z. 2012, ArXiv e-prints
  1212.3063

\bibitem[{{Mar{\'{\i}}n-Franch} {et~al.}(2009){Mar{\'{\i}}n-Franch},
  {Aparicio}, {Piotto}, {Rosenberg}, {Chaboyer}, {Sarajedini}, {Siegel},
  {Anderson}, {Bedin}, {Dotter}, {Hempel}, {King}, {Majewski}, {Milone},
  {Paust}, \& {Reid}}]{2009ApJ...694.1498M}
{Mar{\'{\i}}n-Franch}, A., {Aparicio}, A., {Piotto}, G., {et~al.} 2009, \apj,
  694, 1498

\bibitem[{{McLaughlin} \& van~der Marel(2005)}]{mclaughlin_resolved_2005}
{McLaughlin}, D.~E. \& van~der Marel, R.~P. 2005, \apjs, 161, 304

\bibitem[{Pearson(1901)}]{PCA}
Pearson, K. 1901, Philosophical Magazine, 2 (11), 559

\bibitem[{{Piotto} {et~al.}(2007){Piotto}, {Bedin}, {Anderson}, {King},
  {Cassisi}, {Milone}, {Villanova}, {Pietrinferni}, \&
  {Renzini}}]{2007ApJ...661L..53P}
{Piotto}, G., {Bedin}, L.~R., {Anderson}, J., {et~al.} 2007, \apjl, 661, L53

\bibitem[{{Piotto} {et~al.}(2002){Piotto}, {King}, {Djorgovski}, {Sosin},
  {Zoccali}, {Saviane}, {De Angeli}, {Riello}, {Recio-Blanco}, {Rich},
  {Meylan}, \& {Renzini}}]{2002A&A...391..945P}
{Piotto}, G., {King}, I.~R., {Djorgovski}, S.~G., {et~al.} 2002, \aap, 391, 945

\bibitem[{{Piotto} {et~al.}(2005){Piotto}, {Villanova}, {Bedin}, {Gratton},
  {Cassisi}, {Momany}, {Recio-Blanco}, {Lucatello}, {Anderson}, {King},
  {Pietrinferni}, \& {Carraro}}]{2005ApJ...621..777P}
{Piotto}, G., {Villanova}, S., {Bedin}, L.~R., {et~al.} 2005, \apj, 621, 777

\bibitem[{{Raimondo} {et~al.}(2002){Raimondo}, {Castellani}, {Cassisi},
  {Brocato}, \& {Piotto}}]{2002ApJ...569..975R}
{Raimondo}, G., {Castellani}, V., {Cassisi}, S., {Brocato}, E., \& {Piotto}, G.
  2002, \apj, 569, 975

\bibitem[{Recchi \& Danziger(2005)}]{recchi_self-enrichment_2005}
Recchi, S. \& Danziger, I.~J. 2005, \aap, 436, 145

\bibitem[{Recio-Blanco {et~al.}(2006)Recio-Blanco, Aparicio, Piotto, de~Angeli,
  \& Djorgovski}]{recio-blanco_multivariate_2006}
Recio-Blanco, A., Aparicio, A., Piotto, G., de~Angeli, F., \& Djorgovski, S.~G.
  2006, \aap, 452, 875

\bibitem[{{Reimers}(1977)}]{1977A&A....57..395R}
{Reimers}, D. 1977, \aap, 57, 395

\bibitem[{{Rood}(1973)}]{1973ApJ...184..815R}
{Rood}, R.~T. 1973, \apj, 184, 815

\bibitem[{{Sandage} \& {Wallerstein}(1960)}]{1960ApJ...131..598S}
{Sandage}, A. \& {Wallerstein}, G. 1960, \apj, 131, 598

\bibitem[{{Searle} \& {Zinn}(1978)}]{1978ApJ...225..357S}
{Searle}, L. \& {Zinn}, R. 1978, \apj, 225, 357

\bibitem[{{Spitzer}(1987)}]{1987degc.book.....S}
{Spitzer}, L. 1987, {Dynamical evolution of globular clusters} (Princeton
  University Press, Princeton, NJ)

\bibitem[{{Testa} {et~al.}(2001){Testa}, {Corsi}, {Andreuzzi}, {Iannicola},
  {Marconi}, {Piersimoni}, \& {Buonanno}}]{2001AJ....121..916T}
{Testa}, V., {Corsi}, C.~E., {Andreuzzi}, G., {et~al.} 2001, \aj, 121, 916

\bibitem[{{Trenti} {et~al.}(2010){Trenti}, {Vesperini}, \&
  {Pasquato}}]{2010ApJ...708.1598T}
{Trenti}, M., {Vesperini}, E., \& {Pasquato}, M. 2010, \apj, 708, 1598

\bibitem[{{Yoon} {et~al.}(2008){Yoon}, {Joo}, {Ree}, {Han}, {Kim}, \&
  {Lee}}]{2008ApJ...677.1080Y}
{Yoon}, S.-J., {Joo}, S.-J., {Ree}, C.~H., {et~al.} 2008, \apj, 677, 1080

\end{thebibliography}
\bibliographystyle{aa}

\end{document}